\documentclass[aps,prl,twocolumn,groupedaddress,epsfig,superscriptaddress,showpacs]{revtex4-1}
\usepackage{graphicx}
\usepackage{bm}
\usepackage{CJKutf8}
\usepackage{caption} 
\usepackage{color}
\usepackage{ragged2e}
\renewcommand{\raggedright}{\leftskip=0pt \rightskip=0pt plus 0cm}

\begin{document}


\title{Ultra-high-density local structure of liquid water}


\author{Cheng Yang}
\affiliation{Beijing National Laboratory for Condensed Matter Physics and CAS Key Laboratory of Soft Matter Physics, Institute of Physics, Chinese Academy of Sciences, Beijing 100190, China}
\affiliation{School of Physical Sciences, University of Chinese Academy of Sciences, Beijing 100049, China}

\author{Chuanbiao Zhang}
\affiliation{Department of Physics and Electronic Engineering, Heze University, Heze 274015, China}
\affiliation{School of Physical Sciences, University of Chinese Academy of Sciences, Beijing 100049, China}

\author{Fangfu Ye\footnote{E-Mail: fye@iphy.ac.cn
}}
\affiliation{Beijing National Laboratory for Condensed Matter Physics and CAS Key Laboratory of Soft Matter Physics, Institute of Physics, Chinese Academy of Sciences, Beijing 100190, China}
\affiliation{School of Physical Sciences, University of Chinese Academy of Sciences, Beijing 100049, China}

\author{Xin Zhou\footnote{E-Mail: xzhou@ucas.ac.cn
}}
\affiliation{School of Physical Sciences, University of Chinese Academy of Sciences, Beijing 100049, China}


\date{\today}

\begin{abstract}

The local structure of liquid water plays a key role in determining the anomalous properties of water. We run all-atom simulations for three microscopic water models, and use multiple order parameters to analyse the local structure of water. We identify three types of local structures. In addition to the well known low-density-liquid and high-density-liquid structures, the newly identified third type possesses an ultra high density and overcoordinated H-bonds. The existence of this third type decreases the rate of transition from the high-density-structure to low-density-structure and increases the rate of the reverse one, leading to the enhancement of the high-density-structure stability.

\end{abstract}


\maketitle


Water plays a very important role in daily life and many physical, chemical and biological processes. 
Although being one of the most common substances on earth, it has many unusual thermodynamic properties.
For example, the increases of its isothermal compressibility, isobaric heat capacity and the magnitude of thermal expansion coefficient upon cooling are contrary to normal liquids~\cite{debenedetti2003supercooled}. Clarification of the local structure of liquid water is the key of understanding the origin of these anomalies\cite{russo2014understanding,ponyatovsky1998the,farrell2014clusters}. Furthermore, water molecules interact with each other and form a hydrogen-bond network. The dynamic process of each molecule is confined by the network. Any changes of the position or orientation of individual water molecules influence their neighbors strongly, yielding a collective motion in the hydrogen-bond network. The dynamic inhomogeneity of the network is directly related to the local structures of liquid water \cite{shiratani1996growth,matsumoto2007topological}. 

In this letter, we perform all-atom molecular dynamics simulations and employ multiple order parameters to characterize the local structure of three microscopic water models at temperatures ranging from supercooled to ambient. We identify three types of local structures in liquid water. In addition to the well known high-density-liquid (HDL) and low-density-liquid (LDL) 
\cite{poole1992phase, wernet2004structure, tokushima2008high, huang2009inhomogeneous, paolantoni2009tetrahedral, nilsson2011perspective, nilsson2012fluctuations, pallares2014anomalies, abascal2010widom, kesselring2012nanoscale, liu2012liquid, li2013liquid, giovambattista2013liquid, poole2013free, palmer2014metastable, yagasaki2014spontaneous,liu2007observation,mallamace2007anomalous,sellberg2014ultrafast, Kim1589,xu2005relation}, 
there exists a third type, characterized by an ultra-high density and overcoordinated H-bonds. The concentration of this third type increases when temperature rises. By analyzing the correlation of these three types, we show that the ultra-high-density-liquid (UHDL) molecules  are surrounded by HDL molecules and distribute discretely in space; and we also show that UHDL decreases the rate of transition from the HDL structure to LDL structure and increases the rate of the reverse one, leading to the enhancement of the HDL-structure stability.

The three water models we use are TIP4P/2005~\cite{abascal2005general}, SPC/E~\cite{berendsen1987missing} and TIP4P-Ew~\cite{horn2004development}. All the simulations were performed in isothermal-isobaric(NPT) ensembles, with those of TIP4P/2005 and SPC/E models performed by using the molecular dynamics package LAMMPS~\cite{lammps} and those of TIP4P-Ew by GROMACS~\cite{gromacs}. Long-range solvers were used to compute the long-range coulombic interactions. 

During the simulations, the real structures of molecules were exported, and then minimized by the conjugate gradient algorithm to obtain the inherent structures\cite{debenedetti2001supercooled}, so as to remove the vibrational components of molecules and show the local structure clearly\cite{accordino2011quantitative}. 
In recent studies, such minimisation method has been successfully employed to identify two distinctly different local structures, HDL and LDL~\cite{appignanesi2009evidence, wikfeldt2011spatially}.

To characterize the local structure of liquid water, we focus on two order parameters, the local structure index (LSI)~\cite{wikfeldt2011spatially} and the distance to the fifth closest neighbor ($r_5$)~\cite{cuthbertson2011mixturelike}. The LSI of molecule $i$, $I(i)$, is defined as follows. Assume the distances ($r_j$'s) between molecule $i$ and its neighbours $j$'s can be sorted as $r_1<r_2<r_3\cdots<r_{n(i)}<0.37 \rm nm <r_{n(i)+1}$, where $0.37\rm nm$ sets the starting distance of the second shell; we then have 
\begin{equation}
I(i)=\frac{1}{n(i)}\sum_{j=1}^{n(i)}[\Delta(j,i)-\overline{\Delta(i)}]^2, 
\end{equation}
with
$\Delta(j,i)=r_{j+1}-r_j$ and $\overline{\Delta(i)}$ denoting the average of $\Delta(j,i)$ over all molecules whose distances to molecule $i$ are less than $0.37 \rm nm$. 
And the $r_5$ characterizes the distance between molecule $i$ and its fifth closest neighbor.

\begin{figure}[ht]
\includegraphics[width=0.45\textwidth]{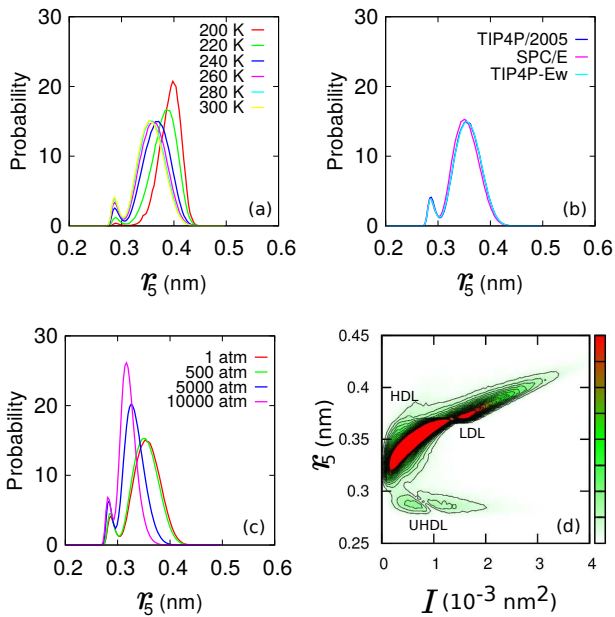}
\caption{\small \raggedright Probability distributions of $r_5$ and combinational $I$-$r_5$ distribution.  (a) $r_5$ distribution of TIP4P/2005 at 1atm for various temperatures, given by simulations of 216 molecules; (b) $r_5$ distribution of TIP4P/2005, TIP4P-Ew and SPC/E models at 1atm and 300K, given by simulations of 1000 molecules; (c) $r_5$ distribution of TIP4P/2005 at 300K and various pressures, given by simulations of 4000 molecules; (d) joint $I$-$r_5$ distribution of TIP4P/2005 at 300K and 1atm (from simulations of  216 molecules), where the values of the probability density are given by the colour bar on the right.}
\label{Fig.1}
\end{figure}     

We first investigate the probability distributions of $r_5$ at various temperatures and pressures.  As shown in Fig.~\ref{Fig.1} (a)-(c), these distributions are bimodal, even under ambient conditions. The position of the left small peaks in these distributions, almost coincident with the first peak position of water's radial distribution function, is smaller than $0.3 \rm nm$, indicating that the fifth neighbour locates inside the first shell. Therefore, such bimodal distributions (under ambient conditions) are different from those previously reported for HDL and LDL~\cite{cuthbertson2011mixturelike}. The height of the left peaks increases when the temperature or pressure increases. The right peak, which is much higher than the left one, is very sensitive to the temperature and/or pressure changes, implying it probably results from the mixing of multiple components. Note that  different microscopic models yield very similar results [Fig.~\ref{Fig.1}(b)].  

To understand the origin of the aforementioned left peak, we proceed to calculate the joint distribution of $r_5$ and LSI, i.e., the $I$-$r_5$ map, at $1$atm and $300$K. Surprisingly, as shown in Fig.~\ref{Fig.1}(d), there now exist three peaks, indicating the presence of three types of local structure of liquid water. The upper right peak, with large LSI and large $r_5$, corresponds to the LDL structure~\cite{wikfeldt2011spatially}. Among the two peaks with small LSI, the upper left one, with larger $r_5$ and much larger height, consists the main part of the low-LSI component~\cite{wikfeldt2011spatially}, and should thus correspond to the HDL structure; the lower left peak, with the smallest height, has the smallest $r_5$, representing a local structure with density even higher than HDL, which we term as ultra-high-density liquid (UHDL) structure~\cite{UhdlBi}. When projected onto the vertical ($r_5$) axis, the HDL and LDL peaks together yield the large peak given in Fig.~\ref{Fig.1}(a)-(c), and the UHDL peak corresponds to the small peak in Fig.~\ref{Fig.1}(a)-(c).  When projected onto the $I$-axis, i.e., the horizontal axis of Fig.~\ref{Fig.1}(d), the UHDL peak is concealed by the HDL and LDL peaks.

\begin{figure}[ht]
\includegraphics[width=0.48\textwidth]{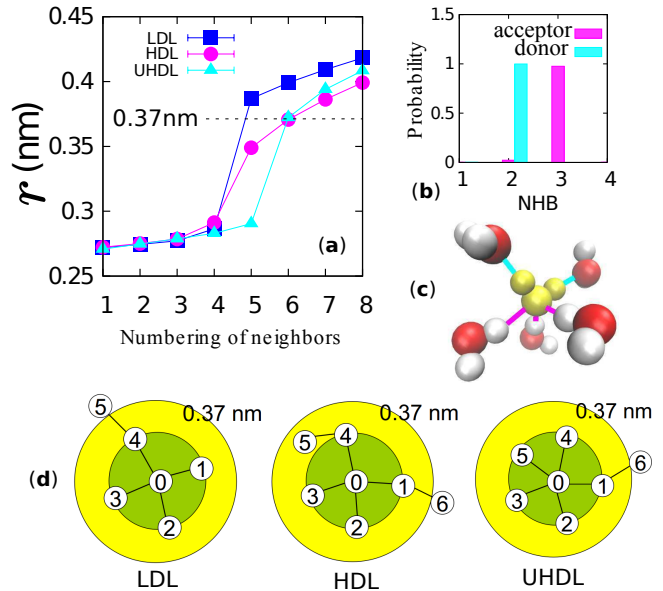}
\caption{\small \raggedright (a) Distance between the center molecule to its neighbours; (b) distributions of UHDL's acceptor and donor hydrogen bonds at 1atm and 300K, given by 216-molecule simulations; (c) snapshot of UHDL, where the cyan and pink lines represent respectively donor and acceptor hydrogen bonds; (d) schematic illustration of three species, where the fifth neighbour of the LDL molecule locates beyond the 0.37nm circle and that of the HDL molecule is in the gap between the first and second shells while the UHDL molecule has five neighbours in the first shell.}
\label{Fig.2}
\end{figure} 

We proceed to investigate the UHDL structure at the molecular level. As shown in Fig.~\ref{Fig.2}(a), in contrast to the HDL and LDL structures, in which the distance of the fifth neighbour to the center molecule is much larger than those of the previous four neighbours, the UHDL structure possesses a first shell containing five molecules, viz., the distance gap now exists between the fifth and sixth molecules rather than the fourth and fifth. Fig.~\ref{Fig.2}(b) gives the hydrogen bond number distribution of UHDL and clearly shows that UHDL has three acceptors and two donors. Fig.~\ref{Fig.2}(c) gives a snapshot of UHDL in the inherent structure. A schematic illustration showing the differences between LDL, HDL, and UHDL is given in Fig.~\ref{Fig.2}(d): (i) LDL is ice-like, with the fifth neighbour locating beyond the 0.37nm cutoff circle; (ii) HDL has four neighbours in the first shell, and the fifth neighbour is in the gap between the first and second shells; (iii) UHDL, however, has five neighbours in the first shell, with no molecule in the gap. The LDL and HDL structures are consistent with the descriptions in Ref.~\cite{debenedetti2003supercooled}. Note that UHDL is different from the very-high-density amorphous ice (VHDA), which has four neighbours in the first shell and two interstitial molecules~\cite{finney2002structure}; it has an over-coordinated structure, but different from the three-donor transition state as reported in Ref.\cite{laage2006molecular}. 

\begin{figure}[ht]
\includegraphics[width=0.45\textwidth]{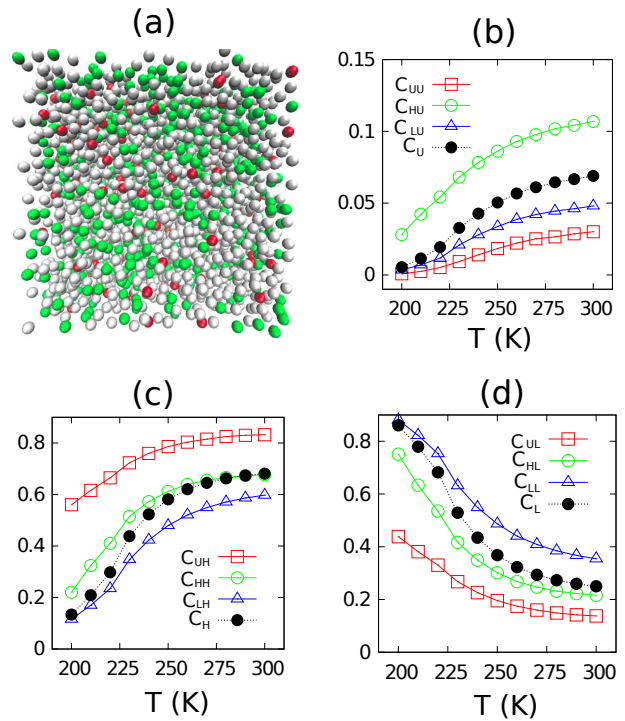}
\caption{\small \raggedright (a) Simulation snapshot of $4000$ liquid-water molecules at $1$atm and $300$K, where the green, white and red spheres represent the LDL, HDL and UHDL molecules, respectively; (b)-(d) temperature dependence of $C_\alpha$ and $C_{\alpha \beta}$ at $1$atm, given by $216$-molecule simulations.}
\label{Fig.3}
\end{figure}

We next discuss the functions of UHDL molecules. To this end, we first analyse the spatial correlations between LDL, HDL, and UHDL molecules. Fig.~\ref{Fig.3}(a) gives a snapshot of liquid water at $300$K and $1$atm, where the green, white, and red spheres represent the oxygen atoms of LDL, HDL, and UHDL molecules, respectively. As shown in Fig.~\ref{Fig.3}(a), the LDL and HDL molecules prefer to clustering with the same species, but the UHDL molecules are dispersed among the HDL. In order to confirm this observation, we calculate the percentage of $\beta$ species in the first-shell neighbours of $\alpha$-species molecules, $C_{\alpha \beta }$, where $\alpha$ and $\beta$ represent LDL, HDL, or UHDL (abbreviated as L, H, and U, respectively), and the concentration of $\alpha$ species in the whole system, $C_\alpha$.  By definition, we have $\sum_{\alpha} C_{\alpha}=1$ and $\sum_{\beta} C_{\alpha \beta}=1$. If $C_{\alpha \beta }$ is larger than $C_\beta$, species $\alpha$ and $\beta$ attract each other; otherwise, $\alpha$ and $\beta$ repel each other. Fig.~\ref{Fig.3}(b)-(d) give the values of the three $C_{\alpha}$'s and six $C_{\alpha \beta}$'s at $1$atm and various temperatures between $200$K and $300$K, from which we can clearly see that HDL molecules attract UHDL molecules while LDL molecules repel UHDL (because $C_{HU}>C_{U}>C_{LU}$). The value of $C_{UU}$ is smaller than $C_{U}$, indicating that UHDL molecules tend not to form clusters. These two results together suggest that UHDL molecules probably serve as the clustering nuclei of HDL molecules. The attraction ``strength" of the nuclei can be characterized by the ratio between $C_{UH}$ and $C_{H}$. As shown in Fig.~S4 in the Supplementary Information, this strength increases when temperature decreases.

\begin{figure}[ht]
\includegraphics[width=0.45\textwidth]{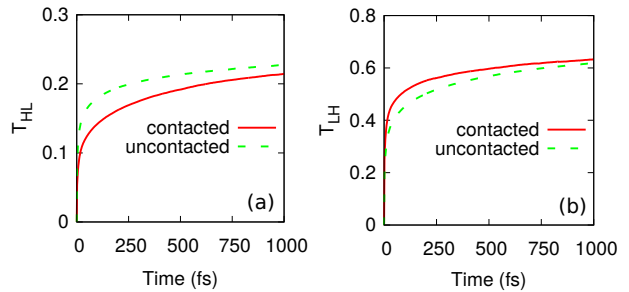}
\caption{\small \raggedright  Transition matrix elements $\rm T_{HL}$ and $\rm T_{LH}$ at 1atm and 300K, given by 4000-molecule simulations, where the red solid lines are for the molecules in the first shell of UHDL (Contacted) and the green dashed lines are for the non-first-shell molecules (Uncontacted).}
\label{Fig.4}
\end{figure}

The existence of UHDL also influences the transitions between the HDL and LDL structures. To quantify such influences, we compare the transition matrix elements, $\rm T_{HL}$ (representing the probability of transition from the HDL structure to the LDL structure) and $\rm T_{LH}$ (from LDL to HDL), of the molecules which are in the first shell of UHDL with those of the molecules that are not. As shown in Fig.~\ref{Fig.4},
the HDL molecules in the first shell of  the UHDL have a lower probability of jumping to LDL, and the first-shell LDL molecules have a higher probability of jumping to HDL, than their non-first-shell couterparts. In another word, the existence of UHDL makes HDL more stable and LDL less stable. 

In conclusion, we used multiple order parameters to describe the local structure of liquid water at various conditions, and found that the local structure of water molecules has three types: low density liquid (LDL), high density liquid (HDL), and ultra-high density liquid (UHDL). The newly-identified UHDL has an over-coordinated structure and possesses the highest local density. It disperses in space and is surrounded by HDL molecules. The existence of UHDL makes HDL more stable and LDL less stable. 
We focused on the inherent structure of liquid water, which was obtained by minimising local potential energies of the real structure. As shown in the Supplemental Material, this minimization process does not change properties of water molecules, and the UHDL can also be observed in the real structure at low temperatures. 

\begin{acknowledgments}
We thank James Farrell for helpful discussions. This work was supported by the Hundred-Talent Program of the Chinese Academy of Sciences (CAS), the Key Research Program of Frontier Sciences of CAS [Grant No. QYZDB-SSW-SYS003],and the National Natural Science Foundation of China (Grants No.11574310 and No.11774394).
\end{acknowledgments}

\bibliographystyle{apsrev4-1}
\bibliography{ref.bib} 

\begin{thebibliography}{40}
\bibitem [{Uhd()}]{UhdlBi}%
  \BibitemOpen
  \href@noop {} {\bibinfo  {journal} {The bimodal characteristic of the UHDL
  peak is caused by the fluctuation of the sixth neighbor around the cutoff at
  $0.37$nm}\ }\BibitemShut {NoStop}%
\end{thebibliography}


\end{document}



\title{Supplemental Material \\
“ Ultra-high-density local structure of liquid water"}


\author{Cheng Yang}
\affiliation{Beijing National Laboratory for Condensed Matter Physics and CAS Key Laboratory of Soft Matter Physics, Institute of Physics, Chinese Academy of Sciences, Beijing 100190, China}
\affiliation{School of Physical Sciences, University of Chinese Academy of Sciences, Beijing 100049, China}

\author{Chuanbiao Zhang}
\affiliation{Department of Physics and Electronic Engineering, Heze University, Heze 274015, China}
\affiliation{School of Physical Sciences, University of Chinese Academy of Sciences, Beijing 100049, China}

\author{Fangfu Ye\footnote{E-Mail: fye@iphy.ac.cn
}}
\affiliation{Beijing National Laboratory for Condensed Matter Physics and CAS Key Laboratory of Soft Matter Physics, Institute of Physics, Chinese Academy of Sciences, Beijing 100190, China}
\affiliation{School of Physical Sciences, University of Chinese Academy of Sciences, Beijing 100049, China}

\author{Xin Zhou\footnote{E-Mail: xzhou@ucas.ac.cn
}}
\affiliation{School of Physical Sciences, University of Chinese Academy of Sciences, Beijing 100049, China}






\maketitle




Minimising local potential energies of the real structure removes the thermal fluctuation of the conformation, yielding the inherent structure. We first compare the distributions of different order parameters (including $r_5$ and $I$) in the real structure (obtained by averaging over many conformations) with those in the inherent structure. Fig.~S1(a)\&(b) show, respectively, the distribuitons of $r_5$ and $I$ of the real structure, which are unimodal except that the distribution of $I$ at low temperatures shows a slight bimodal characteristic. Fig.~S1(c)\&(d) show, respectively, the distributions of $r_5$ and $I$ of the inherent structure, which are bimodal at all temperatures. 
 
\begin{figure}[ht]
\includegraphics[width=0.8\textwidth]{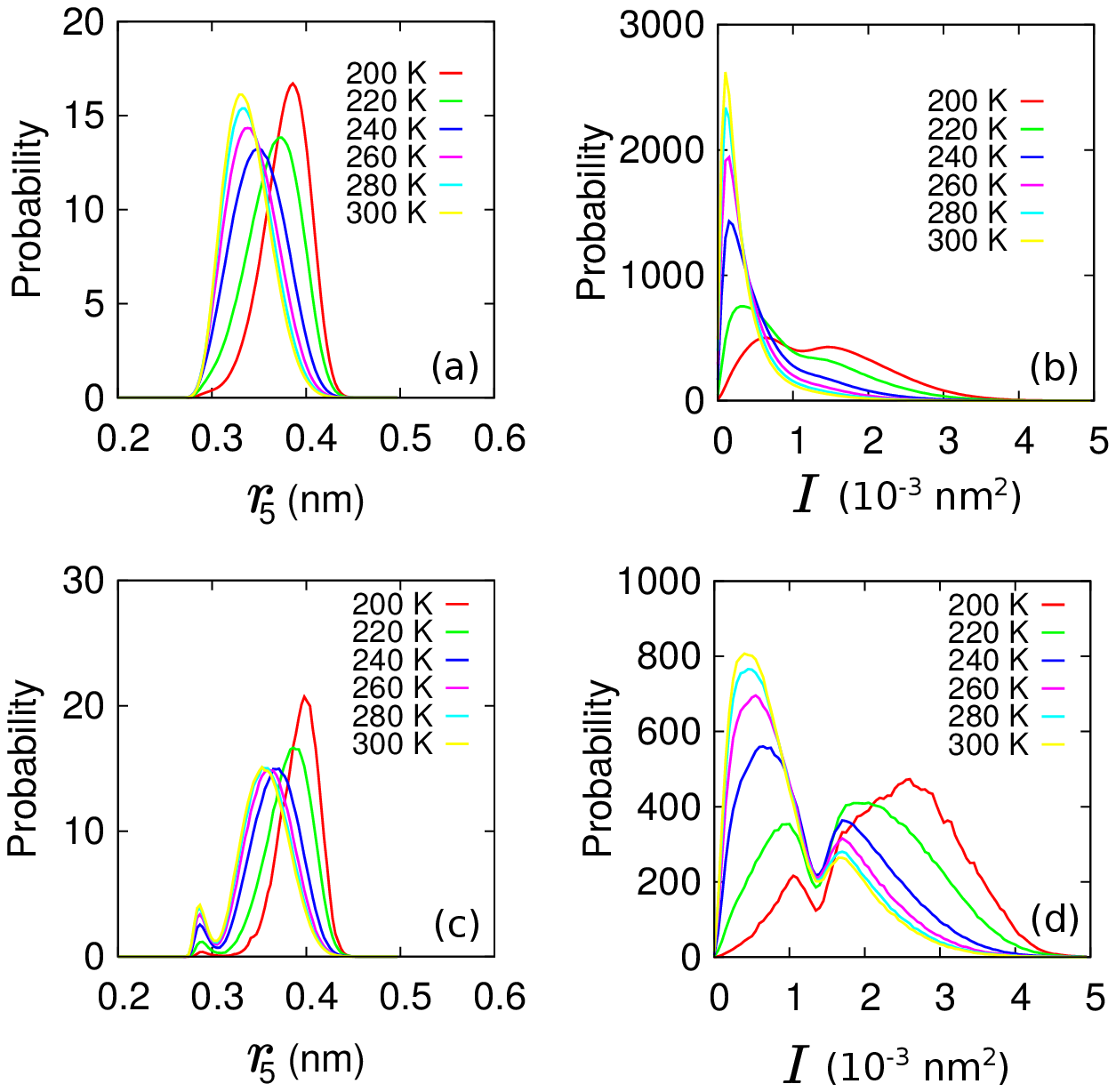}
\caption*{\small Fig.~S1 Distributions of $r_5$ and $I$ at different temperatures in the real and inherent structures: (a) distributions of $r_5$ in the real structure; (b) distributions of $I$ in the real structure; (c) distribuitons of $r_5$ in the inherent structure; (d) distributions of $I$ in the inherent structure.}
\label{s1}
\end{figure} 

The UHDL can also be observed in the real structure at low temperatures. 
To this end, we traced each water molecule to obtain single-molecule trajectories. During the simulation, the water molecule's local structure jumps between different types. When the temperature is very low, this jump does not occur very frequently, and different local structures can be distinguished via proper order parameters. We calculated the probability distribution of $r_5$ of single-molecule trajectories, and all the results are presented in Fig.~S2(a), in which the black thick line is the distribution obtained by averaging all the single-molecule distributions. Although the averaged distribution is unimodal, a small percentage of the single-molecule distributions have bimodal-peaks, as shown in Fig.~S2(b). The position of the left peak in these distributions clearly show the presence of UHDL. 

\begin{figure}[ht]
\includegraphics[width=0.8\textwidth]{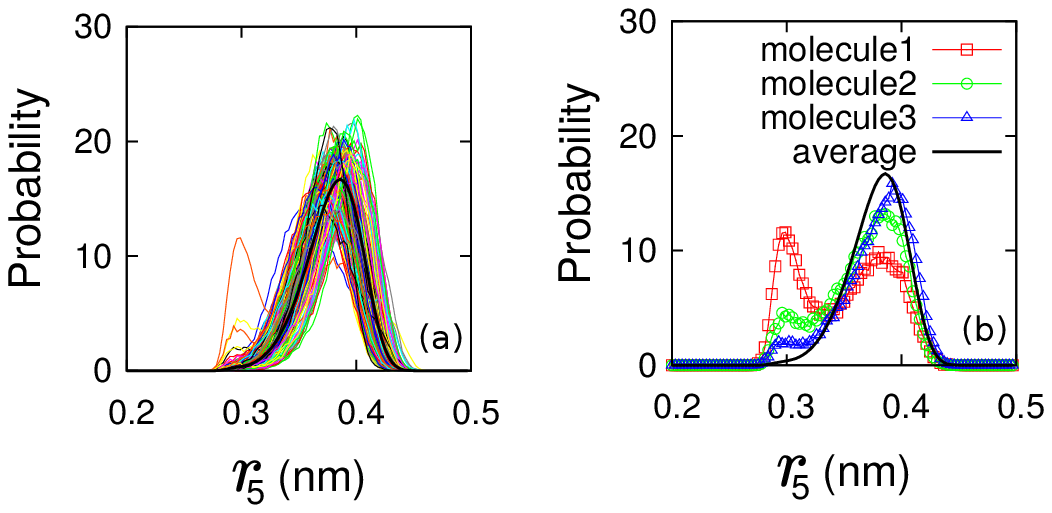}
\caption*{\small Fig.~S2  Distributions of $r_5$ of single-molecule trajectories in the real structure given by 216-molecule simulations at 200K and 1 atm: (a) distributions of all 216 molecules, (b) distributions of three molecule trajectories showing bimodal structure, in which the black thick line is the averaged distribution.}
\label{s2}
\end{figure}

We further investigate whether the minimization process of yielding the inherent structure changes the properties of water molecules. We classify molecules into three species according to their local structures in the inherent structure. We first compute the hydrogen bond numbers and local density (the inverse of the Voronoi polyhedron) of each molecule when they are in the inherent structure, and then compute the corresponding values when they are in the real structure.  The comparison of these sets of values are given in Fig.~S3, in which the upper row shows the distributions of LDL, HDL, and UHDL molecules in the inherent structure and the bottom row shows the corresponding distributions in the real structure. These results in the two rows are almost identical, indicating the minimization process does not obviously change the properties of water molecules.

\begin{figure}[ht]
\includegraphics[width=0.8\textwidth]{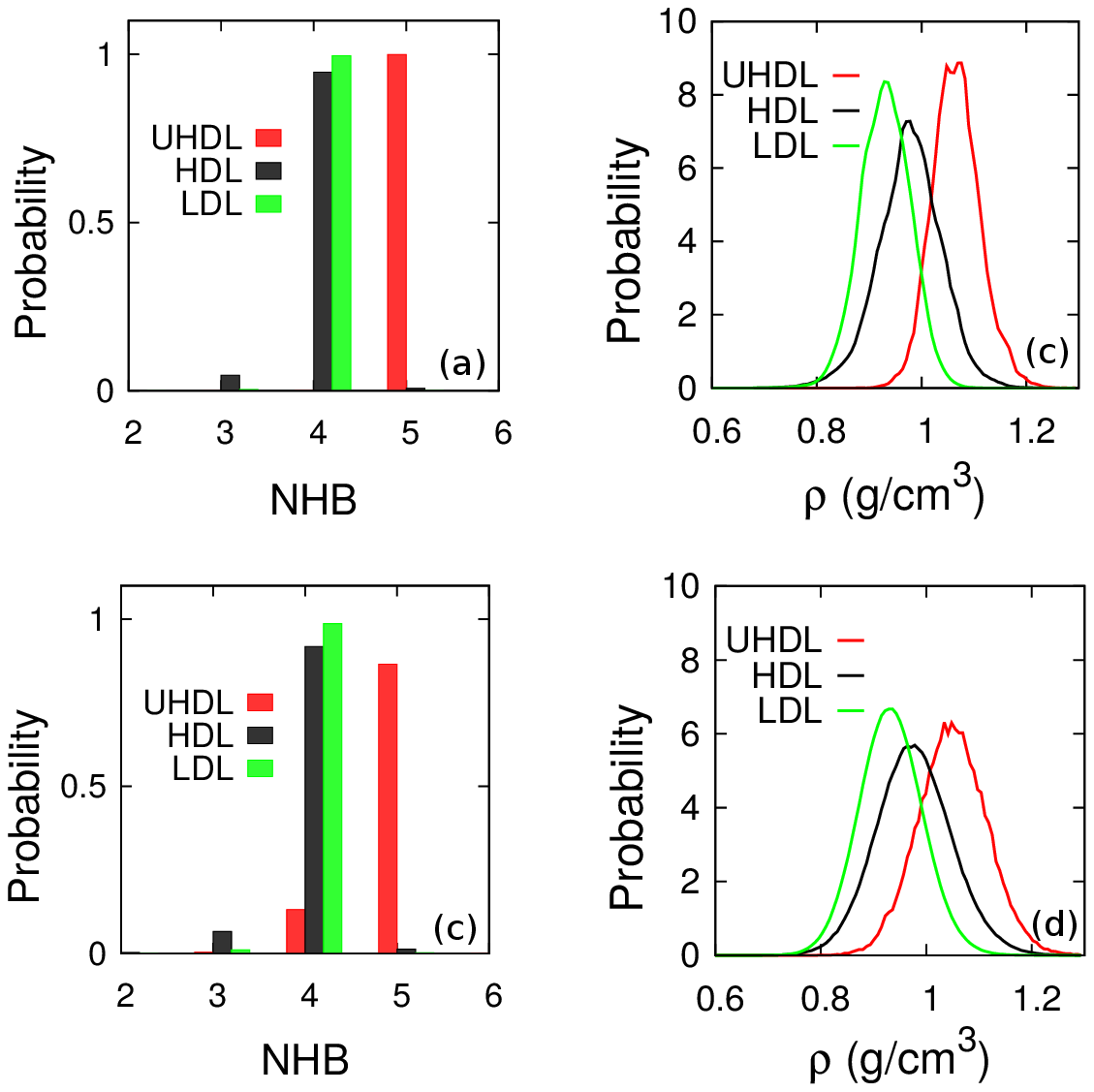}
\caption*{\small Fig.~S3 Distributions of hydrogen-bond numbers and local density of three species in the inherent structure [(a)\&(b)] and the real structure [(c)\&(d)], given by 216-molecule simulations at 200K and 1atm.}
\label{Fig.S3}
\end{figure} 

Fig.~S4 gives the values of various $C'_{\alpha \beta}=C_{\alpha \beta}/C_\beta$, among which $C'_{UH}$ represents the attraction ``strength" of the UHDL molecule to their HDL neighbours.  The value of $C'_\beta$ is unity. If $\alpha$ and $\beta$ attract each other, $C'_{\alpha \beta}$  and $C'_{\beta \alpha}$ will be larger than one; otherwise, they will be smaller than one. 

\begin{figure}[ht]
\includegraphics[width=0.8\textwidth]{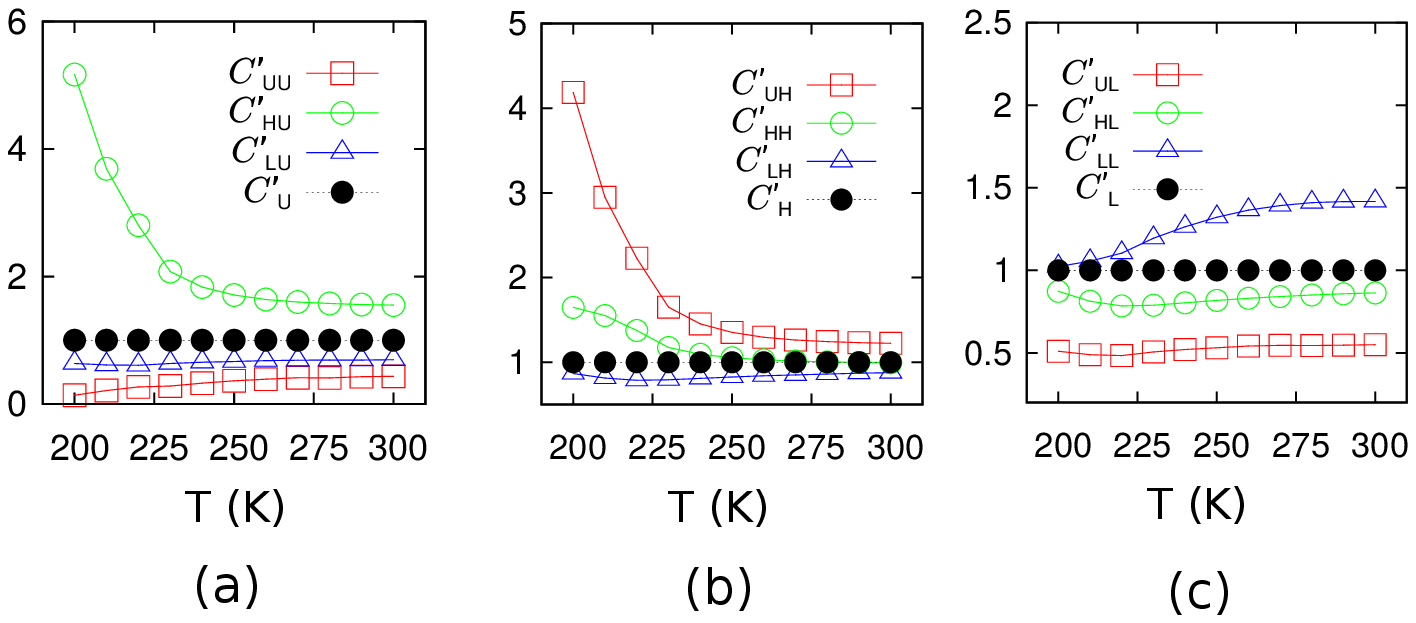}
\caption*{\small Fig.~S4: Ratio between $C_{\alpha\beta}$ and $C_{\beta}$ given by 216-molecule simulations at 1atm and various temperatures.}
\label{Fig.S4}
\end{figure} 

\bibliographystyle{apsrev4-1}